%% file: Paper.tex
\documentclass[lettersize,journal]{IEEEtran}
\pdfoutput=1
\usepackage{cite}
\usepackage{mathtools,amssymb,amsmath,amsthm} 
\usepackage{graphicx} 
\graphicspath{ {Figures/} } 
\usepackage{float}
\usepackage{color}
\DeclareMathOperator{\sgn}{sgn}
\newtheorem{theory}{Theorem}
\usepackage{xcolor}
\usepackage{tikz}
\usetikzlibrary{patterns}
\usepackage{acro} 
\newcommand{\norm}[1]{\left\lVert#1\right\rVert_{1}}
\def\BibTeX{{\rm B\kern-.05em{\sc i\kern-.025em b}\kern-.08em
		T\kern-.1667em\lower.7ex\hbox{E}\kern-.125emX}}

\DeclareAcronym{SINR}{
	short=SINR,
	long=signal-to-interference-plus-noise ratio,
}
\DeclareAcronym{HOL}{
	short=HOL,
	long=head-of-line,
}

\begin{document}
	
	\title{Analysis of Half-Duplex Two-Node Slotted ALOHA \\ Network With Asynchronous Traffic}
\author{Seyed~Ali~Hashemian and
	Farid~Ashtiani,~\IEEEmembership{Senior Member,~IEEE}
	\thanks{The authors are with the Department of Electrical Engineering, Sharif University of Technology (SUT). Tehran 11155-4363, Iran (email: a.hashemian17@sharif.edu, ashtianimt@sharif.edu) }
}
{}	
	\maketitle
	
	\begin{abstract}
Despite the long history of research on slotted ALOHA, the exact analysis of the average delay is still in question as the performance of each node is coupled with the activity of other nodes. In this paper, we consider a network comprised of two half-duplex transmitter nodes with asynchronous arrival traffic that follow the slotted ALOHA protocol. We propose a new queueing theoretic model based on the state-dependent queues to analyze the network. In addition, we derive the exact values of delay and stability region for each node. The numerical results demonstrate the accuracy of our proposed model.
	\end{abstract}
	
	
	\input{"sec01_intro"}
	\input{"sec02_system_model"}
	\input{"sec03_queueing_model"}

	\input{"sec04_stability"}
	\input{"sec05_numerical"}

	\input{"sec06_conclusion"}
	\ifCLASSOPTIONcaptionsoff
	\newpage
	\fi
	
	\bibliographystyle{IEEEtran}
	\bibliography{Bibfiles/IEEEabrv,Bibfiles/references}

\end{document}

%% file: sec01_intro.tex
\section{Introduction}
Despite the popularity of slotted ALOHA and its wide deployment in practical systems, e.g., internet of things (IoT) networks, due to its simplicity and lack of need for centralized coordination, the performance analysis of the ALOHA system has not been done perfectly. Slotted ALOHA is a random access protocol where the channel resources are shared among multiple users. Each user transmits data at time slots if it has packets and backs off when a collision happens according to feedback from the destination \cite{book:Bertsekas}. The increasing number of devices in the fifth and sixth generations of mobile networks (5G and 6G), especially in the case of massive machine-type communications (mMTC) as well as the increasing popularity of IoT technology, has renewed the motivation to use random access schemes as a simple yet effective uplink access protocol in large networks \cite{osseiran2014, Clazzer2019, Munari2019}. 

The analysis of the stability region of the slotted ALOHA network, i.e., the incoming packet rate that can be allowed without causing the network to be saturated, has been a subject of research for a long time. The authors in \cite{Rao1988} considered a system of multiple slotted ALOHA transmitters and a common receiver and derived the exact stability region for the case of two transmitters and the inner bounds of the stability region for the case of multiple transmitters by exploiting the stochastic dominance technique. In \cite{Pappas2013}, the authors considered a two-transmitter two-receiver system with the collision modeled as the probability of the \ac{SINR} at receivers being below a certain threshold. Then, by using similar dominant systems, the authors found the stability region for two types of receivers: one that treats interference as noise and the other that has successive interference cancellation. In \cite{Zhong2016}, the authors considered multiple transmitter-receiver pairs and instead of calculating the strict stability region, defined the $\epsilon$-stability and, using the dominance technique, derived sufficient and necessary conditions for network $\epsilon$-stability. In \cite{moradian2012}, the authors considered a system of slotted ALOHA with multiple energy harvesting transmitters and, considering a queueing network for each \ac{HOL} packet, derived the network stability region by solving the traffic equations in an iterative manner. In \cite{Farhadi2014} the authors considered a slotted ALOHA network with K-exponential back-off and modeled the interactions between nodes using interconnected quasi-birth-death (QBD) processes with a few phases and infinite levels. Thus, the stability region as well as the average delay were derived by solving the QBD processes iteratively.

Despite its importance, the exact calculation of delay is a relatively cumbersome task even in the simplest networks with interactive queues. This is because the success probability of packet transmissions in each queue depends on the state of other queues. In other words, the queues are coupled, and without knowing the joint steady-state distribution of all queues, one cannot separately calculate the service rate for the individual queues. Some works derived a non-closed form or an approximate value for the delay. The authors in \cite{Dai2009A, Dai2009B, Dai2012} considered the network of multiple buffered ALOHA nodes with K-exponential back-off that was modeled as a multi-queue-single-server system and developed an approximate approach to find stable regions based on back-off parameters in addition to the delay value for each packet. In \cite{Dimitriou2018}, the authors considered the cases of two and three transmitting nodes and a single common receiver with multi-packet reception capabilities under a queue-aware transmission policy (i.e., a node is aware of the state of its neighbors). Assuming that the transmitting nodes can change their transmission probabilities based on the status of the other nodes, they provided the stability region using the dominance technique. In addition, for the case of two nodes and using the theory of boundary value problem, they obtained the average queueing delay in the form of expressions for the general case and the explicit forms for the special case of symmetric transmitting nodes and a common receiver with single packet reception. The authors in \cite{bergamo2021} considered the similar two-transmitter two-receiver model with coupled service rates as in \cite{Dimitriou2018} and, using the transition matrix of the Markov chain, approximately derived the steady-state probability, delay, and throughput for a finite-state equivalent system and then discussed the optimum system parameters. The authors in \cite{zhong2015} considered a network of multiple pairs of transmitters and receivers and found the upper and lower bounds for the delay using the dominance technique.

Independent arrivals at different transmitters is a common assumption. However, in some relaying scenarios (e.g.,\cite{Michalopoulos2015, Bletsas2006}), the traffic generated by relays may be asynchronous, meaning that packets cannot arrive at relays simultaneously. In fact, opportunistic relay selection (ORS) is a relaying technique where the source activates one of the available relays through a selection process based on some criteria, e.g., channel state information (CSI) to forward data. The ORS increases diversity, reduces pathloss and shadowing effects, and, at the same time, avoids excessive use of network resources \cite{Nomikos2016}. Then, the relays send their packets in a coordinated manner or independently (e.g., via a random access protocol). The latter scenario motivates us to specifically study nodes with asynchronous arrival traffic in a slotted ALOHA network.

Simultaneous transmission and reception in slotted ALOHA transceivers may be possible (full-duplex) or not (half-duplex). Despite better spectral efficiency, full-duplex systems require complex hardware to cancel the self-interference, making the operation challenging for small-size and economical devices \cite{kumar2019}. Therefore, in this paper, we study a half-duplex system rather than a common full-duplex one. We consider a scenario of two transmitters following a slotted ALOHA protocol transmitting on a shared channel where any simultaneous transmissions result in a collision. We then propose a queueing theoretic model based on the state-dependent queues to analyze the interactions among coupled nodes and derive the steady-state probability distribution of each queue along with the stability region and the exact values for the delay. In summary, our main contributions are
\begin{itemize}
	\item Proposing a state-dependent approach for modeling two coupled queues with asynchronous arrival traffic in a half-duplex slotted ALOHA network and deriving the steady-state probability distribution
	\item Deriving the exact values for the packet delay and the stability region of the network.
\end{itemize}

The rest of the paper is organized as follows. In Section \ref{sec:system}, we present the system model as well as our assumptions. In Section \ref{sec:queueing}, we propose a state-dependent queueing theoretic approach to derive the exact network steady-state probability distribution. Furthermore, in Section \ref{sec:stability}, we use our framework to derive the stability region and the average delay for each queue. In Section \ref{sec:numerical}, in order to validate our model, we compare the analytic and simulation results. Finally, in Section \ref{sec:conclusion}, we provide concluding remarks.

%% file: sec02_system_model.tex
\section{System Model} \label{sec:system}
\begin{figure}[!t]
	\centerline{ \includegraphics[width=\columnwidth]{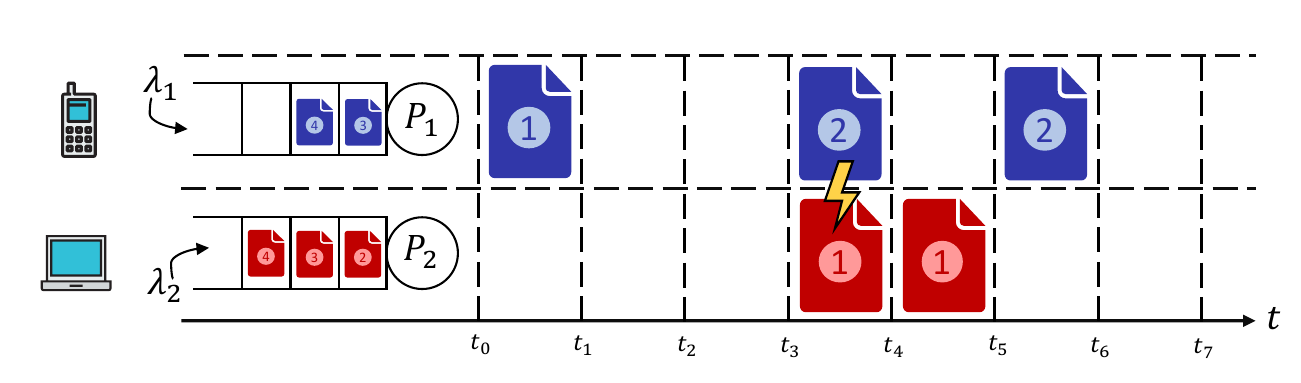}}
	\caption{Packet arrival and transmission. Collision happens at slot 3.}
	\label{fig:sysmodel}
\end{figure}

As shown in Fig \ref{fig:sysmodel}, we consider a network consisting of a common receiver and a set $\mathcal{I} = \{1, 2\}$ of transmitter nodes following the slotted ALOHA protocol. Time is equally slotted and slot $k = \{0,1, \dots\}$ is bounded by $t_{k}$ and $t_{k+1}$. In each time slot $k$, a packet arrives at node $i \in \mathcal{I}$ with probability $\lambda_{i}$. We assume that packets do not arrive at both nodes simultaneously, i.e., arriving data packets are generated by mutually exclusive events.

Each node then transmits arrived packets one by one and in the order of their arrivals (first-come first-serve policy). We assume that it takes a full slot to transmit a packet. This assumption is easily relaxed by considering some error probability at each single transmission due to channel impairments. We also assume that each node has a half-duplex functionality, i.e., at each time slot, the node cannot transmit and receive packets simultaneously. Therefore, in each time slot, node $i \in \mathcal{I}$ transmits a packet with probability $P_{i} < 1$ only if there is no packet arrival in that time slot. In other words, arrivals are privileged. 

In slotted ALOHA, nodes transmit their packets on a common channel, and therefore, when two nodes attempt to transmit packets simultaneously, collision, and thereupon, unsuccessful packet reception happens at the receiver. At the end of each transmission slot, an immediate ACK/NACK feedback from the receiver indicates whether the slot contained a successful transmission or a collision \cite{book:Bertsekas}. In the case of collision in a time slot, nodes must retransmit the collided packets until a successful transmission occurs. Since in each time slot only a single packet departure is possible in the entire network, for the sake of stability, the total arrival rate must be less than one, i.e., $\lambda = \lambda_{1} + \lambda_{2} \leq 1$.

%% file: sec03_queueing_model.tex
\section{Queueing Theoretic Analysis} \label{sec:queueing}
In this section, we propose a queueing theoretic model that can be used to derive the steady-state probability distribution of each node and, therefore, analyze the packet delay and the stability region. 

\subsection{Our Proposed State-Dependent Queueing Model}
Considering packets as customers, each node $i \in \mathcal{I}$ can be modeled as a $Geo/Geo(\bar{n})/1$ discrete-time queue with arrival rate $\lambda_{i}$ and a service rate that depends on the state of other queues as collision can reduce the probability of a packet leaving the queue. This means that we have a network of queues that are coupled in service rate and cannot be analyzed independently using conventional queueing theoretic approaches. More specifically, since the state of each queue depends on the other queue, a closed-form expression for the steady-state distribution cannot be obtained. At first glance, it seems that queues can be solved simultaneously using a two-dimensional Markov chain, but since the number of states is infinite, this approach requires a truncation approximation. In the following, we propose a state-dependent queueing network model to address coupled service rates. Hereafter, we will use the terms node and queue interchangeably. 

Consider a non-negative vector $\bar{n} = (n_1, n_2)\in \bar{N} = \mathbb{Z}^{2}_{\geq 0}$ as the network state vector where $n_1$ and $n_2$ represent the number of customers in queues $1$ and $2$, respectively. Also consider a two-part and non-negative vector $\bar{\bar{a}} = (\bar{a}, \bar{a}^{*}) \in \bar{\bar{A}}$ as the movement vector with $\bar{a} = (a_1, a_2)$ and $\bar{a}^{*} = (a^{*}_1, a^{*}_2)$ where the elements $a_1$ and $a_2$ (resp. $a^{*}_1$ and $a^{*}_2$) represent the number of departures from (resp. arrivals to) queues $1$ and $2$, respectively. Moreover, $\bar{\bar{A}}$ contains all valid movement vectors regarding our system model. We can consider nodes $0_1$ and $0_2$ as the exogenous world corresponding to nodes $1$ and $2$ such that departures from $0_i$, correspond to arrivals at node $i$ and vice versa.

Let $\bar{n} \in \bar{N}$ be the network state at the beginning of a typical time slot and let the movement $\bar{\bar{a}} \in \bar{\bar{A}}$ be realized during that time slot. Vector $\bar{\bar{a}}$, then, transforms into another vector $\bar{\bar{a}}' \in \bar{\bar{A}}$ with probability $r((\bar{n}, \bar{\bar{a}}), (\bar{n}',\bar{\bar{a}}'))$ which changes the network state to $\bar{n}' = \bar{n} - \bar{a} + \bar{a}'$. We define $\langle \bar{n}, \bar{\bar{a}} \rangle$ as the set of all pairs of $(\bar{n}, \bar{\bar{a}})$ that can transform into each other, i.e., $\bar{n} - \bar{a} = \bar{n}' - \bar{a}'$ and $\norm{\bar{\bar{a}}} = \norm{\bar{\bar{a}}'}$.

It follows from our assumptions in Section \ref{sec:system} that at each time slot, only a single departure and a single arrival are possible, but they cannot happen at the same node. Therefore, all possible movements in a time slot can be described as $\bar{\bar{a}}(i, j) =  (\bar{a}(i), \bar{a}^{*}(j)) \in \bar{\bar{A}}, i \neq j$ where $i, j \in \{0, 1, 2\}$. The index $i$ (resp. $j$) represents the node with a departure (resp. arrival) and determines which element of $\bar{a}$ (resp. $\bar{a}^{*}$) equals one. Value $0$ refers to the outside of the system and $\bar{\bar{a}}(0, 0)$ is an all-zero vector that describes the case when there are no movements in a time slot. As a result, we have all possible movements $\bar{\bar{a}} \in \bar{\bar{A}}$ as
\begin{equation}
	\begin{gathered}
		\bar{\bar{a}}(0, 1) = (0, 0, 1, 0), \hspace{0.5cm} \bar{\bar{a}}(1, 0) = (1, 0, 0, 0),\\
		\bar{\bar{a}}(0, 2) = (0, 0, 0, 1), \hspace{0.5cm} \bar{\bar{a}}(2, 0) = (0, 1, 0, 0),\\
		\bar{\bar{a}}(1, 2) = (1, 0, 0, 1), \hspace{0.5cm} \bar{\bar{a}}(2, 1) = (0, 1, 1, 0), \\
		\bar{\bar{a}}(0, 0) = (0, 0, 0, 0),
	\end{gathered}
\end{equation}
and $\exists \bar{n}, \bar{n}' \in \bar{N}$ where
\begin{equation}
	\begin{gathered}
		\langle \bar{n}, \bar{\bar{a}}(0, 1) \rangle = \langle \bar{n}', \bar{\bar{a}}(1, 0) \rangle; \hspace{2em} \bar{n}'\geq \bar{a}(1),\\
		\langle \bar{n}, \bar{\bar{a}}(0, 2) \rangle = \langle \bar{n}', \bar{\bar{a}}(2, 0) \rangle; \hspace{2em} \bar{n}'\geq \bar{a}(2),\\
		\langle \bar{n}, \bar{\bar{a}}(1, 2) \rangle = \langle \bar{n}', \bar{\bar{a}}(2, 1) \rangle; \hspace{2em} \bar{n}\geq \bar{a}(1), \bar{n}'\geq \bar{a}(2),\\
		\langle \bar{n}, \bar{\bar{a}}(0, 0) \rangle = \{(\bar{n}, \bar{\bar{a}}(0, 0))\}.
	\end{gathered}
\end{equation}

\subsection{The Network Steady-State Probability Distribution}
Let us define $\kappa_{r}$ as the state space of the Markov chain defined by transition function $r$ where
\begin{equation}
	\kappa_{r}\! =\!  \{ (\bar{n}, \bar{\bar{a}}) \! \in \! \bar{N} \times \bar{\bar{A}}| \exists (\bar{n}', \bar{\bar{a}}') \!\in \!\bar{N} \times \bar{\bar{A}},  r((\bar{n}', \bar{\bar{a}}'),(\bar{n}, \bar{\bar{a}})) > 0 \}.
\end{equation}
For each possible pair of vectors $(\bar{n}, \bar{\bar{a}}) \in \kappa_{r}$, we define $\xi(\bar{n}, \bar{\bar{a}})$ as the rate of leaving the state $\bar{n}$ through the movement $\bar{\bar{a}}$. Therefore, for $\bar{n} \in \bar{N}$ we can write
\begin{equation} \label{eq:765983}
	\begin{split}
		\xi(\bar{n}, \bar{\bar{a}}(0,1)) &=  \lambda_{1} (1-P_2)^{\sgn(n_2)},\\
		\xi(\bar{n}, \bar{\bar{a}}(1,0)) &= (1-\lambda)P_1(1-P_2)^{\sgn(n_2)}; \hspace{1em} \bar{n}\geq \bar{a}(1),\\
		\xi(\bar{n}, \bar{\bar{a}}(0,2)) &=  \lambda_{2}(1 - P_1)^{\sgn(n_1)},\\
		\xi(\bar{n}, \bar{\bar{a}}(2,0)) &= (1-\lambda)(1-P_1)^{\sgn(n_1)}P_2; \hspace{1em} \bar{n}\geq \bar{a}(2),\\
		\xi(\bar{n}, \bar{\bar{a}}(1,2)) &= \lambda_{2}P_1; \hspace{1em} \bar{n}\geq \bar{a}(1),\\
		\xi(\bar{n}, \bar{\bar{a}}(2,1)) &= \lambda_{1}P_2; \hspace{1em} \bar{n}\geq \bar{a}(2),\\
		\xi(\bar{n}, \bar{\bar{a}}(0,0)) &= (1-\lambda)(1-P_1)^{\sgn(n_1)}(1-P_2)^{\sgn(n_2)} \\&+ (1-\lambda)P_{1}P_{2}\sgn(n_1)\sgn(n_2),
	\end{split}
\end{equation}
where $\sgn(.)$ equals one if its argument is not equal to zero. Note that the assumption that in our half-duplex scenario arrivals have priority over transmissions is reflected in \eqref{eq:765983}. We also define $\pi(\bar{n})$ as the network steady-state probability distribution at the time slot boundaries. Now we use the following theorem from \cite{book:Miyazawa}.
\begin{theory}\label{theory:951737}
Suppose that there exists a non-negative function $\Psi$ on $\kappa_{r}$ and a positive function $\Phi$ on $\bar{N}$ such that $\xi(\bar{n}, \bar{\bar{a}})$ is given by 
\begin{equation} \label{eq:708166}
	\xi(\bar{n}, \bar{\bar{a}}) = \frac{ \Psi(\langle\bar{n}, \bar{\bar{a}}\rangle) \nu(\bar{n}, \bar{\bar{a}})}{\Phi(\bar{n})}; \hspace{0.5cm} (\bar{n}, \bar{\bar{a}}) \in \kappa_{r},
\end{equation}	
where $\nu$ is a stationary measure of $r$ so that for all $(\bar{n}, \bar{\bar{a}}) \in \kappa_{r}$ we have
\begin{equation} \label{eq:660677}
	\nu(\bar{n}, \bar{\bar{a}}) = \sum_{(\bar{n}', \bar{\bar{a}}') \in \kappa_{r}}  \nu(\bar{n}', \bar{\bar{a}}') r((\bar{n}', \bar{\bar{a}}'),(\bar{n}, \bar{\bar{a}})) >0,
\end{equation}	 	
and there exists $c$ such that
\begin{equation} \label{eq:979051}
	c^{-1} = \sum_{(\bar{n}, \bar{\bar{a}}) \in \kappa_{r}}  \Psi(\langle \bar{n}, \bar{\bar{a}} \rangle) \nu(\bar{n}, \bar{\bar{a}}) < \infty.
\end{equation}	
Then, the stationary distribution $\pi(\bar{n})$ is given by
\begin{equation}
	\pi(\bar{n}) = c \Phi(\bar{n}); \hspace{2em} \bar{n} \in \bar{N}.
\end{equation}	
\end{theory}
\begin{proof}
	Please see chapter 9 in \cite{book:Miyazawa}.
\end{proof}

It is followed from Theorem \ref{theory:951737} that if we find functions $\Phi$, $\Psi$ and $\nu$ such that \eqref{eq:708166} and \eqref{eq:979051} are satisfied, then we can derive the $\pi(\bar{n})$. We define $\Psi \triangleq \Psi_{1} \circ \Psi_{2}$ where $\Psi_{2}(\langle \bar{n}, \bar{\bar{a}} \rangle) = \bar{n} - \bar{a}$. For a variable vector $\bar{x} = (x_1, x_2)$, we propose the following functions
\begin{equation} \label{eq:186608}
	\Psi_{1}(\bar{x}) = \Phi(\bar{x}) =  P_{1}^{-x_{1}}P_{2}^{-x_{2}} (\frac{\lambda_{1}}{1-\lambda})^{x_{1}} (\frac{\lambda_{2}}{1-\lambda})^{x_{2}}.
\end{equation}
Also, since in our model, at each time slot only a single packet arrival and departure is possible for each node (no batch movements), for $\{(\bar{n}, \bar{\bar{a}}), (\bar{n}', \bar{\bar{a}}')\} \in \langle\bar{n}, \bar{\bar{a}}\rangle$, we have $r((\bar{n}, \bar{\bar{a}}), (\bar{n}', \bar{\bar{a}}')) = r((\bar{n}', \bar{\bar{a}}'), (\bar{n}, \bar{\bar{a}})) = 1$ and, as a result, $\nu(\langle\bar{n}, \bar{\bar{a}}\rangle) = \nu(\bar{n}, \bar{\bar{a}}) = \nu(\bar{n}', \bar{\bar{a}}')$. Therefore, we propose
\begin{equation} \label{eq:480497}
	\small
	\begin{split}
		\nu(\langle\bar{n}, \bar{\bar{a}}(0,1)\rangle) &=  \lambda_{1} (1-P_2)^{\sgn(n_2)},\\
		\nu(\langle\bar{n}, \bar{\bar{a}}(0,2)\rangle) &= \lambda_{2}(1 - P_1)^{\sgn(n_1)},\\
		\nu(\langle\bar{n}, \bar{\bar{a}}(1,2)\rangle) &= \frac{\lambda_{1}\lambda_{2}}{1-\lambda}, \\
		\nu(\langle\bar{n}, \bar{\bar{a}}(0,0)\rangle) &= (1-\lambda)(1-P_1)^{\sgn(n_1)}(1-P_2)^{\sgn(n_2)} \\&+ (1-\lambda)P_{1}P_{2}\sgn(n_1)\sgn(n_2).
	\end{split}
\end{equation}

Substituting rates from \eqref{eq:765983} and proposed functions from \eqref{eq:186608} and \eqref{eq:480497}, one can verify that condition \eqref{eq:708166} is satisfied.
Therefore, we can derive the network steady-state probability distribution as $\pi(\bar{n}) =  c \Phi(\bar{n})$.

%% file: sec04_stability.tex
\section{Stability Region and Average Delay} \label{sec:stability}
The derived steady-state probability distribution in Section \ref{sec:queueing} is valid for some arrival rates corresponding to our assumptions. In this section, we try to find the boundaries for the arrival rate of each queue. Since we assumed that in each time slot, only a single packet arrival is possible in the entire network, we have
\begin{equation} \label{eq:305660}
	\lambda_{1} + \lambda_{2} \leq 1.
\end{equation}

Besides, from Theorem \ref{theory:951737} and since $\sum_{\bar{n} \in \bar{N}} \pi(\bar{n}) = 1$, we need to have
\begin{equation}
	\sum_{\bar{n}} \Phi(\bar{n}) = c^{-1} < \infty.
\end{equation}
Therefore, we have 
\begin{equation}
	\label{eq:687464}
	\begin{split}
			\sum_{\bar{n}} \Phi(\bar{n}) &= \bigg( \sum_{n_1 = 0}^{\infty} P_{1}^{-n_{1}} (\frac{\lambda_{1}}{1-\lambda})^{n_{1}} \bigg) \bigg(\sum_{n_2 = 0}^{\infty} P_{2}^{-n_{2}}  (\frac{\lambda_{2}}{1-\lambda})^{n_{2}} \bigg)\\
			& = \bigg(\sum_{n_1 = 0}^{\infty}  (\frac{\lambda_{1}}{P_{1} (1-\lambda)})^{n_{1}} \bigg) \bigg(\sum_{n_2 = 0}^{\infty} (\frac{\lambda_{2}}{P_{2} (1-\lambda)})^{n_{2}} \bigg)\\
			& = (1 - \frac{\lambda_{1}}{P_{1}(1- \lambda)})^{-1}(1 - \frac{\lambda_{2}}{P_{2}(1- \lambda)})^{-1} = c^{-1}.
	\end{split}
\end{equation}
In order for \eqref{eq:687464} to converge, for $i \in \mathcal{I}$, we need to have
\begin{equation} \label{eq:208969}
	\frac{\lambda_{i}}{P_{i} (1-\lambda)} < 1 .
\end{equation}

\begin{figure}[!t]
	\centering
	\begin{tikzpicture}[scale=4.5, blacknode/.style={pattern=north west lines, pattern color=magenta, shape=rectangle, minimum width=0.4cm, draw=black, line width=0.5},
		bluenode/.style={shape=rectangle, minimum width=0.4cm, draw=black, dashed, fill=gray!30, fill opacity=0.4, line width=0.5}]
		\draw[->,very thick] (0,0)--(1.1,0) node[right]{$\lambda_1$};
		\draw[->, very thick] (0,0)--(0,1.1) node[above]{$\lambda_2$};
		\draw[ thick, color=red] (0,1)--(0.4,0);
		\draw[ thick, color=blue] (0,0.4)--(1,0);
		\draw[pattern=north west lines, pattern color=magenta] (0,0.4)--(2/7,2/7)--(0.4,0)--(0,0)--(0,0.4);

		\node at (1,-0.04) {\footnotesize $1$};
		\node at (-0.04, 1) {\footnotesize $1$};
		\node at (2/7 + 0.33 ,2/7+0.04) {$ (\frac{P_{1}}{1 + P_{1} + P_{2}}, \frac{P_{2}}{1 + P_{1} + P_{2}})$};
		\node at (0.4 ,0 - 0.06) {$ (\frac{P_{1}}{1 + P_{1}}, 0)$};
		\node at (0 - 0.15, 0.4) {$ (0, \frac{P_{2}}{1 + P_{2}})$};
		\draw[color=black, dashed, fill=gray!30, fill opacity=0.4 ] (0,0)--(2/3,0)--(2/9,2/9)--(0, 2/3)--(0,0);
		\node at (2/3+0.01 ,0 - 0.06) { $(P_{1}, 0)$};
		\node at (0 - 0.14, 2/3) {$(0,P_{2})$};
		\node at (0.5 +0.05 ,0.5+0.04) {\footnotesize $ (P_{1}(1 - P_{2}), P_{2}(1 - P_{1}))$};

		\draw[->, thick] (2/9,2/9)--(0.3,0.45+0.04);
		\matrix [draw,below left] at (1.1,1.1) {
				\node [blacknode,label=right:\text{\footnotesize Our Stability Region}] {}; \\
				\node [bluenode,label=right:\text{\footnotesize Stability Region from \cite{Rao1988}} ] {}; \\
		};
	\end{tikzpicture}
	\caption{Stability region. While the stability region in our work is always convex, the stability region from \cite{Rao1988} may be concave or convex depending on $P_1$ and $P_2$ values.}
	\label{fig:stability_region}
\end{figure}
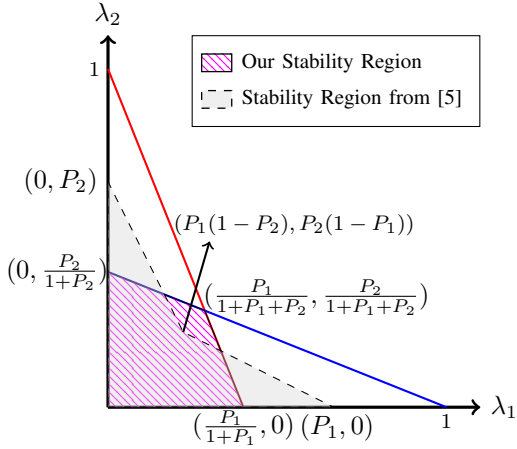

Therefore, \eqref{eq:305660} and \eqref{eq:208969} determine the stability region of the half-duplex network, which is shown in Fig. \ref{fig:stability_region}. For the comparison, we have also drawn the stability region of a two-node full-duplex slotted ALOHA network from \cite{Rao1988}. The area under the stability region for our work and \cite{Rao1988} can be easily calculated as $\frac{P_{1}P_{2}(2 + P_{1} + P_{2})}{2(1+P_{1})(1+P_{2})(1+P_{1} + P_{2})}$ and $\frac{P_{1}P_{2}}{2}(2 - P_{1}-P_{2})$, respectively.

Having the value of $c$ we can achieve the steady-state probability distribution for the network, therefore, for queue $i \in \mathcal{I}$ we have
\begin{equation}
	\begin{split}
		\pi_{i} (n_{i}) = \sum_{n_{-i} = 0}^{\infty} \pi(\bar{n}) &= (1 - \frac{\lambda_{i}}{P_{i}(1- \lambda)})  (\frac{\lambda_{i}}{P_{i} (1-\lambda)})^{n_{i}} \\
		&= (1 - \rho_{i})\rho_{i}^{n_{i}},
	\end{split}
\end{equation}
where $\rho_{i} = \frac{\lambda_{i}}{P_{i}(1- \lambda)}$ and notation $-i$ corresponds to the other node in the network. Having the steady-state probability distribution for each queue, one can derive the average packet delay $D_{i}$ using Little's law as
\begin{equation}
	D_{i} = \frac{\sum_{n_{i} = 0}^{\infty}n_{i}\pi_{i} (n_{i})}{\lambda_{i}} = \frac{\rho_{i}}{\lambda_{i}(1-\rho_{i})}.
\end{equation}

%% file: sec05_numerical.tex
\section{Numerical Results} \label{sec:numerical}
In this section, we provide the numerical results to validate our proposed state-dependent analytical approach. To show the accuracy of our analysis, we compare the analytical results with the simulation results in terms of the stability region and the average delay. We use \textsc{Matlab} environment to achieve both analytical and simulation results. The simulation results are obtained for 200000 time slots.

\begin{figure}[!t]
	\centerline{ \includegraphics[width=0.95\columnwidth]{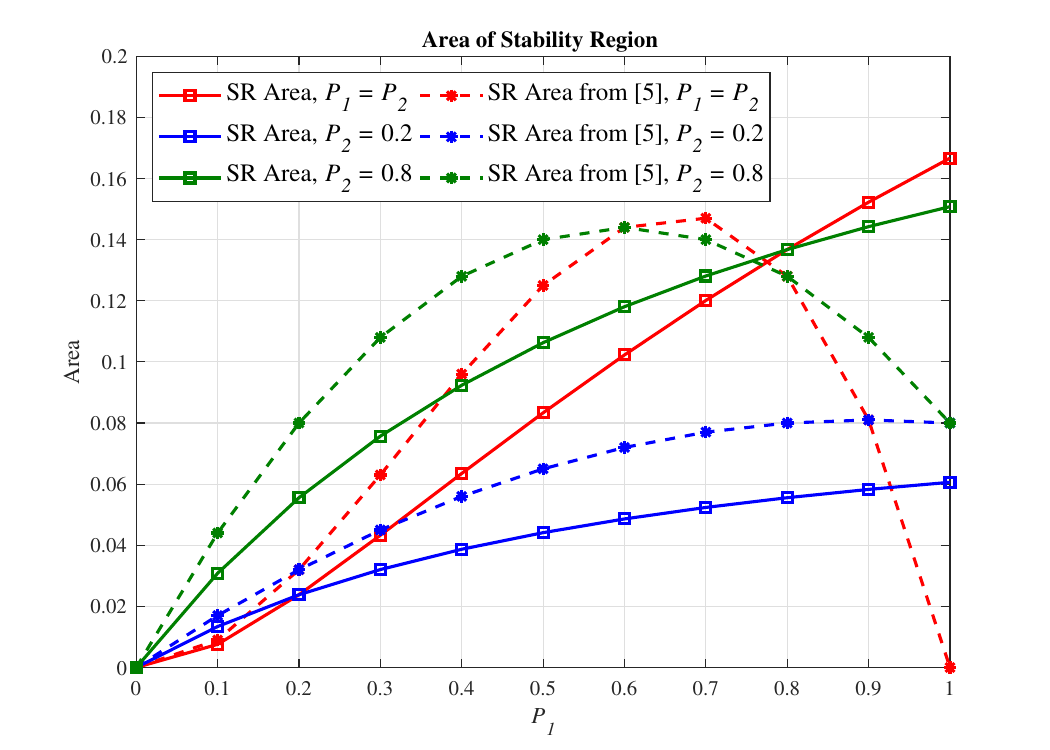}}
	\caption{Area under the stability region for multiple values of $P_1$ and $P_2$}
	\label{fig:area}
\end{figure}

Fig. \ref{fig:stability} represents the boundary of the stability region for multiple values of $P_1$ and $P_2$. To obtain the simulation results, we considered the ratio of the number of arriving packets to the number of successfully transmitted packets over a large time interval. In the case of stability, the ratio is equal to one. For each $\lambda_1$, we found a minimum $\lambda_2$ value that results in a ratio higher than one, indicating that the network is unstable. Moreover, Fig. \ref{fig:area} shows the area under the stability region for multiple values of $P_1$ and $P_2$, as presented in Section \ref{sec:stability}. For the sake of comparison, we have also drawn the area under the stability region for a full-duplex system from \cite{Rao1988}. It can be observed that, at low transmission probabilities, the half-duplex system has a stability region with a smaller area compared to the full-duplex system. This is because the service rate for a half-duplex system is lower than a full-duplex system as no transmission is allowed when there is an arrival. When the transmission probabilities are low, the interference limitation on the stability region is minute compared to the inherent limitation of the half-duplex system. However, for high values of the transmission probabilities, the stability region for a full-duplex system decreases as each node is highly affected by the interference from the other node, i.e., collision. In contrast, the stability region for the half-duplex system increases as the limitation on the transmission limits the interference (collision) as well.
\begin{figure*}[!t]
	\setcounter{figure}{4}
	\centerline{ \includegraphics[width=1.85\columnwidth]{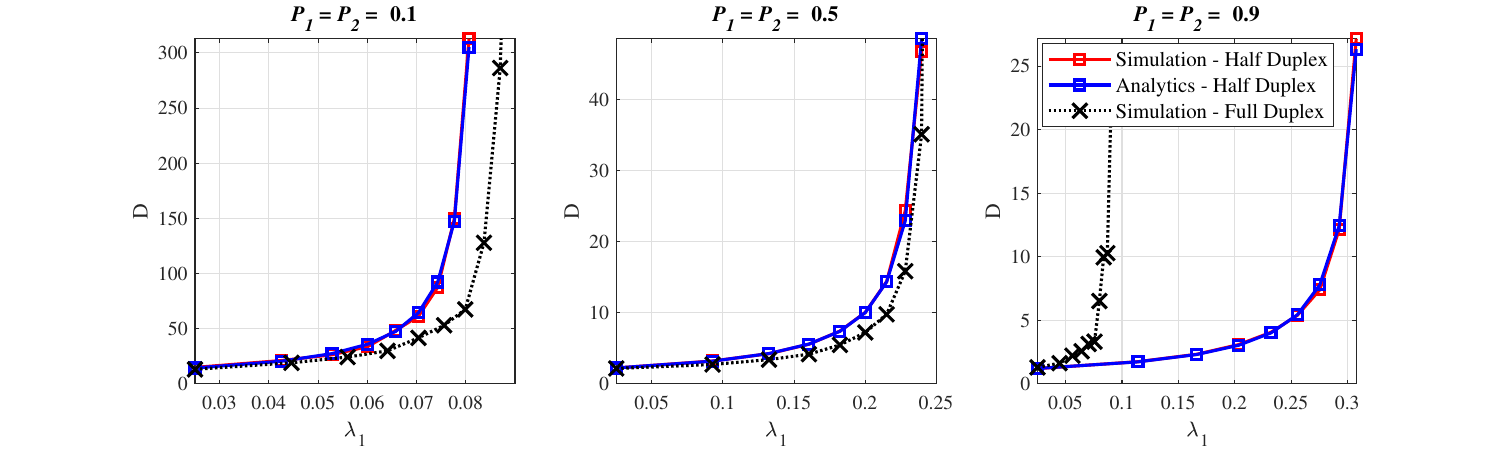}}
	\caption{Average delay (in slots) for a typical packet when $P_1 = P_2$}
	\label{fig:delay}
\end{figure*}
\begin{figure}[!t]
	\setcounter{figure}{2}
	\centerline{ \includegraphics[width=0.95\columnwidth]{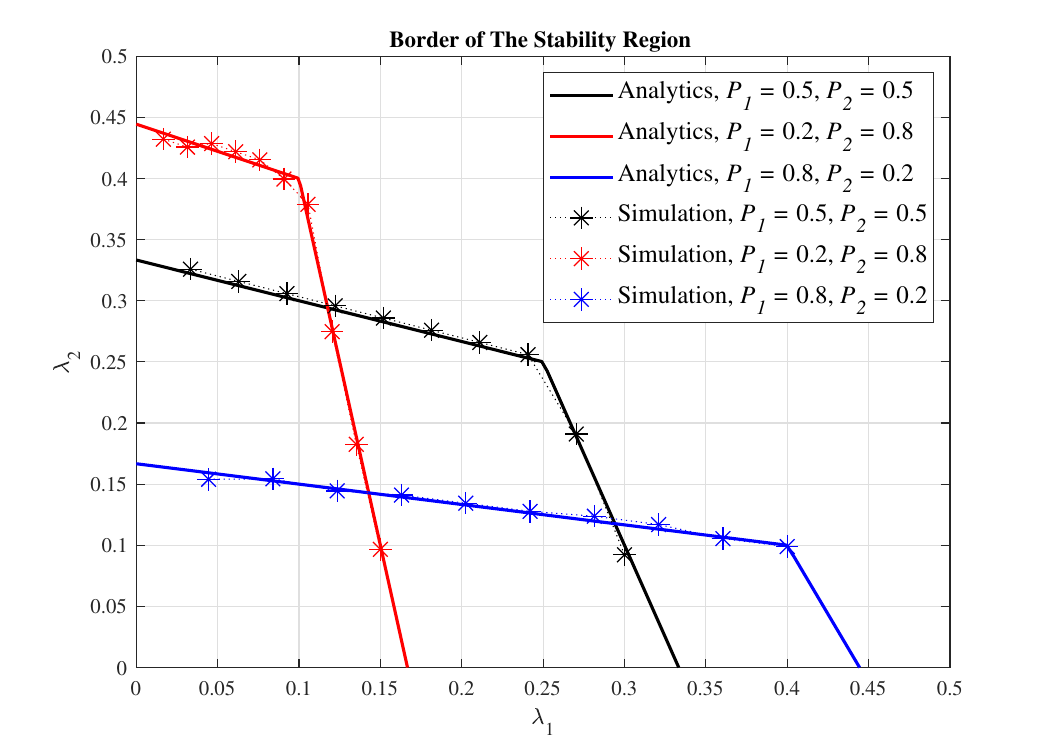}}
	\caption{Stability region for multiple values of $P_1$ and $P_2$}
	\label{fig:stability}
\end{figure}

Next, we consider a symmetric scenario with $P_{1} = P_{2}$ and $\lambda_{1} = \lambda_{2}$ to investigate the delay when the nodes are working within the stability region. Fig. \ref{fig:delay} shows the average packet delay, $D$ (in terms of time slots), versus the arrival rate for multiple values of transmission probabilities. It is obvious that our proposed state-dependent model is accurate and exactly matches with the simulation results. For comparison, we have also drawn the simulation results for a full-duplex system following an early-departure-late-arrival model with independent arrivals at each node. With a similar argument as for the stability region, the delay for a full-duplex system has lower (resp. higher) values for low (resp. high) transmission probabilities compared to the half-duplex system.

%% file: sec06_conclusion.tex
\section{Conclusion} \label{sec:conclusion}
In this work, we investigated a scenario of two half-duplex transmitter nodes with asynchronous arrival traffic following a slotted ALOHA protocol where simultaneous transmissions lead to a collision. We proposed an analytical model comprised of a network of state-dependent queues to achieve the network steady-state probability distribution. Then, for each node, we derived the exact values of the average delay and the stability region. Our numerical results demonstrated that the proposed queueing theoretic model matches with the simulation results. Extending the analytical approach to the full-duplex scenario as well as the case with more nodes is within our future work. 